\documentclass{midl} 

\usepackage{mwe} 
\usepackage{siunitx}
\usepackage{bm}
\usepackage{caption}
\jmlrproceedings{MIDL}{Submitted to MIDL 2020}
\jmlrpages{}

\title[Intervertebral disc detection with redundant counting network]{Spine intervertebral disc labeling using a fully convolutional redundant counting model}

\midlauthor{\Name{Lucas Rouhier\nametag{$^{1}$}} \Email{lucasrouhier@gmail.com}\\
\Name{Francisco Perdigon Romero\nametag{$^{2}$}} \Email{francisco.perdigon@polymtl.ca}\\
\Name{Joseph Paul Cohen\nametag{$^{3}$}} \Email{joseph@josephpcohen.com}\\
\Name{Julien Cohen-Adad\nametag{$^{1,4}$}} \Email{jcohen@polymtl.ca}\\
\addr $^{1}$ NeuroPoly Lab, Institute of Biomedical Engineering, Polytechnique Montreal, Montreal, QC, Canada \\
\addr $^{2}$ MedICAL Laboratory, Polytechnique Montreal, Montreal, QC, Canada\\
\addr $^{3}$ Mila, Universite de Montreal, QC, Canada\\
\addr $^{4}$ Functional Neuroimaging Unit, CRIUGM, Universite de Montreal, Montreal, QC, Canada\\
}

\begin{document}

\maketitle

\begin{abstract}
Labeling intervertebral discs is relevant as it notably enables clinicians to understand the relationship between a patient's symptoms (pain, paralysis) and the exact level of spinal cord injury. However manually labeling those discs is a tedious and user-biased task which would benefit from automated methods. While some automated methods already exist for MRI and CT-scan, they are either not publicly available, or fail to generalize across various imaging contrasts. In this paper we combine a Fully Convolutional Network (FCN) with inception modules to localize and label intervertebral discs. We demonstrate a proof-of-concept application in a publicly-available multi-center and multi-contrast MRI database (n=235 subjects). The code is publicly available at https://github.com/neuropoly/vertebral-labeling-deep-learning. 

\end{abstract}


\section{Introduction}
Detection and labeling of intervertebral discs is useful in a clinical and academic setting to observe the progression of diseases, or to inform analyses in functional MRI results. Numerous automated detection methods were created to achieve this task. Some are based on template matching, which detects the C2/C3 disc with a HOG-SVM model \cite{Gros2017-qd} and then finds the following discs with a sliding window that compares with a probabilistic human spine template \cite{Ullmann2014-lp}. Another method is based on a 3D Fully Convolutional Network (FCN) \cite{Chen2019-ic} that segments the disc and retrieves its center coordinates. However, these methods are sensitive to the variability of MR quality, contrast and resolution. The goals of this study were to (i) adapt an FCN which was shown to work on multimodal CT images for disc segmentation and localization \cite{Chen2019-ic}, (ii) combine the FCN with inception modules to localize intervertebral discs from MRI data and (iii) train the architecture using a publicly-available multi-center and multi-contrast dataset, to strenghten the generalization capabilities of the model. 

\section{Material \& method}

\subsection{Data}
We used the Spinal Cord MRI Public Database \cite{Cohen-Adad2019-vv}. This MRI dataset is composed of T2w and T1w data from 235 subjects, acquired at 40 different centers, thereby exhibiting "real-world" variability in terms of image quality. An average of the 6 middle slices of each subject was used as input images to the network. Ground truths were manually-created by defining a single pixel at the posterior tip of each intervertebral disc. The dataset was split into 75\%, 10\% and 15\% for training, testing, and validation.

\subsection{Preprocessing}
3D volumes were preprocessed using the Spinal Cord Toolbox (SCT) v4.0.1 \cite{De_Leener2016-fg}. They were resampled at 1 mm isotropic resolution and straightened according to the spinal cord centerline \cite{DeLeener2017-tp} obtained with the spinal cord segmentation \cite{Gros2019-ka}. As part of straightening transformation, the image was cropped to 141x141 pixels around the spinal region. A Contrast Limited Adaptive Histogram Equalization algorithm was applied to reduce contrast variability in the image \cite{Zuiderveld1994-lk}. To deal with class imbalance, we increased the target size by applying a 10-pixel Gaussian kernel to single-pixel labels.

\subsection{Processing}
Our custom deep learning model based on inception modules \cite{Szegedy2015-jj} is shown in Figure 1. It extracts several patches within each image, every pixel is therefore processed by the network several times, allowing the model to average over the error and minimize false negatives and false positives, as it was done for counting cells in microscopic slices \cite{Cohen2017-rx}. We trained the network for 1,000 epochs with a combination of dice loss \cite{Milletari2016-hy}, adaptive wing loss \cite{Wang2019-tb} and L2-loss (squared loss). 

\begin{figure}[h]
\includegraphics[width=0.9\linewidth]{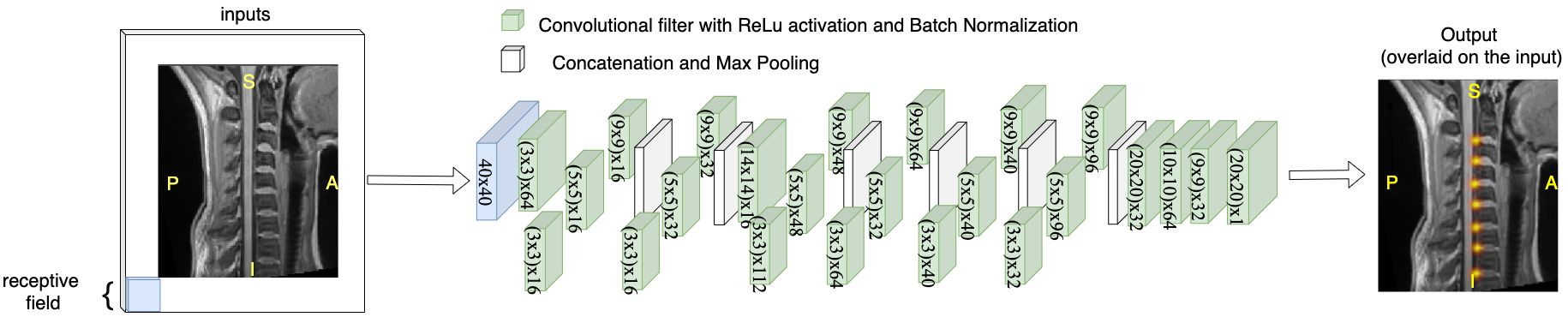}
\caption{\small This shows the model with its receptive fields (40x40) on the input. The input is the sagittal view of
  a T1w MRI and is 0-padded with the size of the receptive field to avoid edge effect. The numbers on the layer represent (kernel size x kernel size) x Number of channels. The output represents the predicted Gaussian functions overlaid on the input from which we will extract discs coordinates.}
\end{figure}

\subsection{Metrics}
Predicted Gaussian functions were thresholded at 0.5 and the center of mass was retrieved as the predicted coordinates. The performance was evaluated based on the distance between the manually-labeled and the predicted coordinates along the superior-inferior axis as well as False positive rate (FPR) and False Negative Rate (FNR). False positives were defined as predicted points that were at least 5 mm away from any ground truth points or groups of predicted points associated with the same ground truth coordinate. False negatives were counted with ground truth points 5 mm or more away from the predicted points.

\section{Results}
Figure 2 compares our results on the validation set with the previous SCT method using template matching \cite{Ullmann2014-lp}, the ablation study (use of a similar neural network built with inception modules without redundant counting) and the architecture with L1-loss. The proposed model works equally well on the two (T1w and T2w) contrasts, improves prediction precision and reduces the number of FNR and FPR on both modalities. All metrics have been computed based on these methods performance on the validation set. 
\begin{figure}[htbp]
 {\includegraphics[width=1\linewidth]{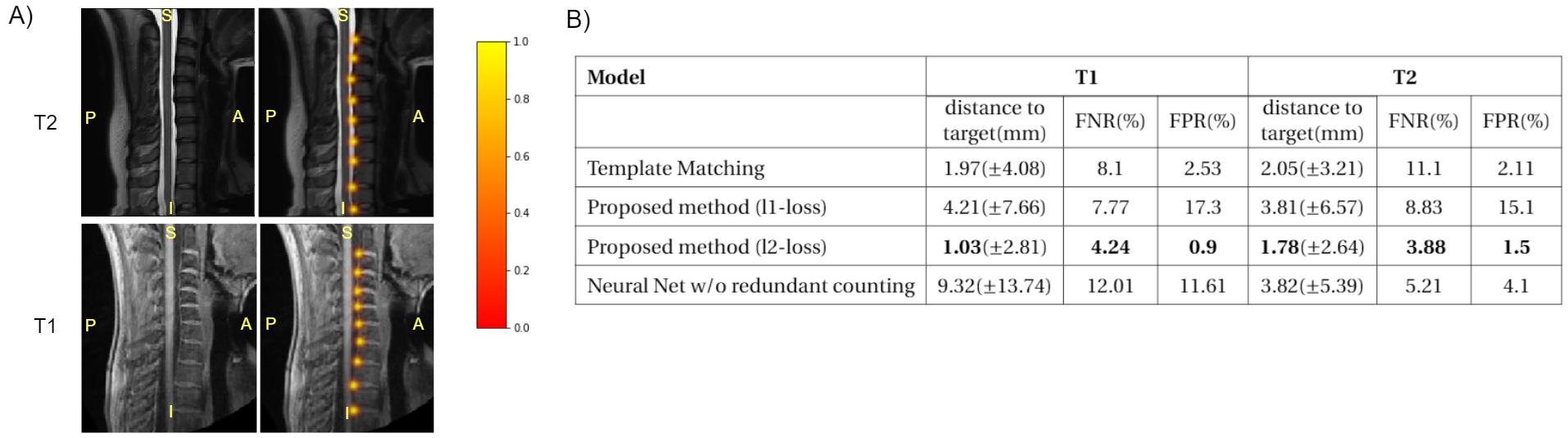}}
  {\caption{\small A) Input images (left) and output predictions of our proposed method (right), color-coded between 0 (transparent/red) and 1 (yellow).\\ 
B) Comparison between SCT’s template-matching method, the proposed method with L1 and L2 loss, as well as inception network without redundant counting on the evaluation metrics. Smaller distance means better precision. The FNR represents the number of false negatives divided by the total number of ground truth points. The FPR is the number of false positives divided by the total number of predicted points. For both ratio, the smaller the better.  
}}
\end{figure}
\section{Conclusion and discussion}
This study presents a new architecture for detecting intervertebral discs. The method shows improvement in precision of localization and decrease of false positives/negatives. Future work will include extending the testing of this model to more "real-life" datasets in patients with spinal pathologies (spinal cord injury, multiple sclerosis, tumors, etc.).

\bibliography{bibliography.bib}

\begin{thebibliography}{12}
\providecommand{\natexlab}[1]{#1}
\providecommand{\url}[1]{\texttt{#1}}
\expandafter\ifx\csname urlstyle\endcsname\relax
  \providecommand{\doi}[1]{doi: #1}\else
  \providecommand{\doi}{doi: \begingroup \urlstyle{rm}\Url}\fi

\bibitem[Chen et~al.(2019)Chen, Gao, Li, Zhao, and Zhao]{Chen2019-ic}
Yizhi Chen, Yunhe Gao, Kang Li, Liang Zhao, and Jun Zhao.
\newblock Vertebrae identification and localization utilizing fully
  convolutional networks and a hidden markov model.
\newblock \emph{IEEE Trans. Med. Imaging}, July 2019.

\bibitem[Cohen et~al.(2017)Cohen, Boucher, Glastonbury, Lo, and
  Bengio]{Cohen2017-rx}
Joseph~Paul Cohen, Genevieve Boucher, Craig~A. Glastonbury, Henry~Z. Lo, and
  Yoshua Bengio.
\newblock {Count-ception: Counting by Fully Convolutional Redundant Counting}.
\newblock In \emph{International Conference on Computer Vision Workshop on
  BioImage Computing}, 2017.

\bibitem[Cohen-Adad(2019)]{Cohen-Adad2019-vv}
J~Cohen-Adad.
\newblock Spinal cord {MRI} public database (multi-subjects), July 2019.

\bibitem[De~Leener et~al.(2016)De~Leener, L{\'e}vy, Dupont, Fonov, Stikov,
  Louis~Collins, Callot, and Cohen-Adad]{De_Leener2016-fg}
Benjamin De~Leener, Simon L{\'e}vy, Sara~M Dupont, Vladimir~S Fonov, Nikola
  Stikov, D~Louis~Collins, Virginie Callot, and Julien Cohen-Adad.
\newblock {SCT}: Spinal cord toolbox, an open-source software for processing
  spinal cord {MRI} data.
\newblock \emph{Neuroimage}, October 2016.

\bibitem[De~Leener et~al.(2017)De~Leener, Mangeat, Dupont, Martin, Callot,
  Stikov, Fehlings, and Cohen-Adad]{DeLeener2017-tp}
Benjamin De~Leener, Gabriel Mangeat, Sara Dupont, Allan~R Martin, Virginie
  Callot, Nikola Stikov, Michael~G Fehlings, and Julien Cohen-Adad.
\newblock Topologically preserving straightening of spinal cord {MRI}, 2017.

\bibitem[Gros et~al.(2017)Gros, De~Leener, Dupont, Martin, Fehlings, Bakshi,
  Tummala, Auclair, McLaren, Callot, and {Others}]{Gros2017-qd}
Charley Gros, Benjamin De~Leener, Sara~M Dupont, Allan~R Martin, Michael~G
  Fehlings, Rohit Bakshi, Subhash Tummala, Vincent Auclair, Donald~G McLaren,
  Virginie Callot, and {Others}.
\newblock Automatic spinal cord localization, robust to {MRI} contrasts using
  global curve optimization.
\newblock \emph{Med. Image Anal.}, 2017.

\bibitem[Gros et~al.(2019)Gros, De~Leener, Badji, Maranzano, Eden, Dupont,
  Talbott, Zhuoquiong, Liu, Granberg, Ouellette, Tachibana, Hori, Kamiya,
  Chougar, Stawiarz, Hillert, Bannier, Kerbrat, Edan, Labauge, Callot,
  Pelletier, Audoin, Rasoanandrianina, Brisset, Valsasina, Rocca, Filippi,
  Bakshi, Tauhid, Prados, Yiannakas, Kearney, Ciccarelli, Smith, Treaba,
  Mainero, Lefeuvre, Reich, Nair, Auclair, McLaren, Martin, Fehlings, Vahdat,
  Khatibi, Doyon, Shepherd, Charlson, Narayanan, and Cohen-Adad]{Gros2019-ka}
Charley Gros, Benjamin De~Leener, Atef Badji, Josefina Maranzano, Dominique
  Eden, Sara~M Dupont, Jason Talbott, Ren Zhuoquiong, Yaou Liu, Tobias
  Granberg, Russell Ouellette, Yasuhiko Tachibana, Masaaki Hori, Kouhei Kamiya,
  Lydia Chougar, Leszek Stawiarz, Jan Hillert, Elise Bannier, Anne Kerbrat,
  Gilles Edan, Pierre Labauge, Virginie Callot, Jean Pelletier, Bertrand
  Audoin, Henitsoa Rasoanandrianina, Jean-Christophe Brisset, Paola Valsasina,
  Maria~A Rocca, Massimo Filippi, Rohit Bakshi, Shahamat Tauhid, Ferran Prados,
  Marios Yiannakas, Hugh Kearney, Olga Ciccarelli, Seth Smith,
  Constantina~Andrada Treaba, Caterina Mainero, Jennifer Lefeuvre, Daniel~S
  Reich, Govind Nair, Vincent Auclair, Donald~G McLaren, Allan~R Martin,
  Michael~G Fehlings, Shahabeddin Vahdat, Ali Khatibi, Julien Doyon, Timothy
  Shepherd, Erik Charlson, Sridar Narayanan, and Julien Cohen-Adad.
\newblock Automatic segmentation of the spinal cord and intramedullary multiple
  sclerosis lesions with convolutional neural networks.
\newblock \emph{Neuroimage}, 184:\penalty0 901--915, January 2019.

\bibitem[Milletari et~al.(2016)Milletari, Ahmadi, Kroll, Plate, Rozanski,
  Maiostre, Levin, Dietrich, Ertl-Wagner, B{\"o}tzel, and
  Navab]{Milletari2016-hy}
Fausto Milletari, Seyed-Ahmad Ahmadi, Christine Kroll, Annika Plate, Verena
  Rozanski, Juliana Maiostre, Johannes Levin, Olaf Dietrich, Birgit
  Ertl-Wagner, Kai B{\"o}tzel, and Nassir Navab.
\newblock {Hough-CNN}: Deep learning for segmentation of deep brain regions in
  {MRI} and ultrasound.
\newblock January 2016.

\bibitem[Szegedy et~al.(2015)Szegedy, Liu, Jia, Sermanet, Reed, Anguelov,
  Erhan, Vanhoucke, and Rabinovich]{Szegedy2015-jj}
Christian Szegedy, Wei Liu, Yangqing Jia, Pierre Sermanet, Scott Reed, Dragomir
  Anguelov, Dumitru Erhan, Vincent Vanhoucke, and Andrew Rabinovich.
\newblock Going deeper with convolutions.
\newblock In \emph{Proceedings of the {IEEE} conference on computer vision and
  pattern recognition}, pages 1--9, 2015.

\bibitem[Ullmann et~al.(2014)Ullmann, Pelletier~Paquette, Thong, and
  Cohen-Adad]{Ullmann2014-lp}
Eug{\'e}nie Ullmann, Jean~Fran{\c c}ois Pelletier~Paquette, William~E Thong,
  and Julien Cohen-Adad.
\newblock Automatic labeling of vertebral levels using a robust template-based
  approach.
\newblock \emph{Int. J. Biomed. Imaging}, 2014:\penalty0 719520, July 2014.

\bibitem[Wang et~al.(2019)Wang, Bo, and Fuxin]{Wang2019-tb}
Xinyao Wang, Liefeng Bo, and Li~Fuxin.
\newblock Adaptive wing loss for robust face alignment via heatmap regression.
\newblock April 2019.

\bibitem[Zuiderveld(1994)]{Zuiderveld1994-lk}
Karel Zuiderveld.
\newblock Contrast limited adaptive histogram equalization.
\newblock In \emph{Graphics gems {IV}}, pages 474--485. Academic Press
  Professional, Inc., August 1994.

\end{thebibliography}

\end{document}